# Time-Aware Models for Software Effort Estimation

Michael Franklin Bosu [1], Stephen G. MacDonell [2], Peter Whigham[3]
[1]*Centre for Information Technology, Waikato Institute of Technology*
[2,3]*Department of Information Science, University of Otago, New Zealand*
[2]stephen.macdonell@otago.ac.nz, [3]peter.whigham@otago.ac.nz

**Abstract**

*It seems logical to assert that the dynamic nature of software engineering practice would mean that software effort estimation (SEE) modelling should take into account project start and completion dates. That is, we should build models for future projects based only on data from completed projects; and we should prefer data from recent similar projects over data from older similar projects. Research in SEE modelling generally ignores these recommendations. In this study two different model development approaches that take project timing into account are applied to two publicly available datasets and the outcomes are compared to those drawn from three baseline (non-time-aware) models. Our results indicate: that it is feasible to build accurate effort estimation models using project timing information; that the models differ from those built without considering time, in terms of the parameters included and their weightings; and that there is no statistical significance difference as to which of the two model building approaches is superior in terms of accuracy.*

## I. INTRODUCTION

Contemporary research efforts to address software effort estimation (SEE) typically develop and evaluate models using one, sometimes more, random split(s) of a secondary dataset of project observations into training and testing sets. Models are built using the training set and model accuracy is assessed on the testing set. In practice, however, organizations accumulate data over time as projects are worked on and are (hopefully) completed. It could be expected, then, that this accumulating data set would be the 'training set', used to build models to estimate the effort of *future* projects as each new project is proposed. Thus we have a disconnect between research and practice. Most effort estimation models developed by the research community disregard project start and/or completion dates [1]; as a result, data from 'future' projects can be used to build predictive models of effort for projects that occurred before them in time. To some extent this may be due to the absence of the necessary time-oriented features in the datasets [1]; however, even for datasets that include timing information, this is widely ignored in SEE research as only two (ISBSG and Finnish datasets) of the six datasets in the public domain with timing information have so far been used in developing software effort estimation models that considers time. To the best of our knowledge, this is the first study to develop time-aware effort estimation models using the NASA93 and Desharnais datasets. In these datasets, the completion dates of projects represent the timing information.

This study therefore explicitly considers the year of project completion and uses only data from completed projects to develop models to estimate the effort of projects completed in subsequent years. Two time-aware approaches; Time-Aware Sequential Accumulation (TASA) and Time-Aware Moving Window (TAMW) are used in model development (see section III for details).

The performance of these time-aware models are then assessed in an absolute sense and in a relative sense against three baseline 'models' – leave-one-out, mean and median.

To the best of our knowledge, this study differs from all previous effort estimation time-aware studies as this study applies the TAMW approach and considers the stability of the models. Our research questions are expressed as follows:

RQ1: Is it feasible to develop accurate effort estimation models using project completion dates?

RQ2: Are the parameters and coefficients of time-aware models stable or volatile?

RQ3: Which of the two time-aware modelling approaches, if either, is superior in terms of accuracy?

The rest of this paper is organized as follows. Section II presents the related work, our research method is presented in section III, in section IV we present our results, section V reports threats to the validity of our study, and section VI comprises a discussion and draws conclusions.

## II. RELATED WORK

Though numerous SEE models have been proposed (see [2]) the number of studies that have considered project timing information in effort estimation is negligible and attributed to very few researchers. This section summarizes the few studies that are related to this research.

Lokan and Mendes [3] applied a moving window of the most recently completed projects to new projects in their effort estimation studies. Their results indicated that use of a moving window of the most recently completed projects contributed significantly to the accuracy of models. In a recent study, Amasaki and Lokan [4] proposed a method that is able to select whether to build a model based on time or to use the growing portfolio of projects. MacDonell and Shepperd [5] applied two timing methods – sequential accumulation of project data over time and constant moving window of size 5 – on a proprietary dataset and obtained improved results over managers' estimates, especially for the moving window approach compared to a LOO approach.

This paper applies the two approaches used by MacDonell and Shepperd [5] to two publicly available datasets, except that the moving window approach, presented in the next section, is dynamic as compared to the fixed window size used in [5] and in earlier similar studies.

## III. RESEARCH METHOD

**Data Grouping.** For each of the two datasets used in this study, an attempt is first made to work with the entire dataset before consideration is given to splitting the data into homogeneous subsets with a view to developing models for each partition. The division of datasets into homogeneous subsets is intended to enable us to identify whether specific partitions of the data exhibit trends that are different from those evident for other partitions, or across the entire dataset. Partitions are typically based on factors such as the type of application, the application domain of the project, and/or the unit or department responsible for development.

Partitions such as these are formed by relying on the visualization of boxplots and the use of Mann-Whitney tests to assess whether observations belong to the same distribution. In this study, data that fell outside the boxplot whiskers of distributions were considered as outliers and were not used in model building. A significance level of 0.05 is used for the Mann-Whitney tests, so groupings that have a p-value greater than 0.05 are taken to belong to the same distribution. Use of these partitions will ensure that models are developed for datasets that as far as possible share similar characteristics.

### A. Datasets

**NASA93 Dataset.** This dataset was collected by NASA and it comprises 93 projects undertaken between 1971 and 1987 (as downloaded from the PROMISE Repository http://openscience.us/repo/). The dataset is structured according to the Constructive Cost Model (COCOMO81) developed by Barry Boehm [6]. It comprises 24 attributes of which 15 are the mandatory effort multipliers.

Preliminary analysis indicated that, due to the diversity of the NASA93 projects, it was neither feasible nor sensible to build time-aware models for the entire dataset, and as such the dataset was split into four subsets. These four subsets are: NASA82, comprising projects developed in 1982 and beyond; Center 2 (C2) and Center 5 (C5) subsets, comprising projects developed at NASA's Center 2 and Center 5, respectively; and Semidetached (SD), which includes projects of the semidetached development mode. Due to space limitations the boxplots are not shown, but they can be found at this link[1]. In addition to outliers being evident in the boxplots, three other projects with atypical characteristics were also not used – two projects with size values greater than their effort values, and a project with a productivity rate (i.e., effort divided by size) more than twice as high as that for the project with the next highest productivity rate, and almost eleven times the mean productivity rate.

**Desharnais Dataset.** The Desharnais dataset was collected by Jean-Marc Desharnais from ten organizations in Canada [7]. The projects in this dataset were undertaken between 1983 and 1988. The dataset consists of 81 records and twelve attributes, including size measured in function points and effort measured in person-hours. We used the version comprising of 77 projects as has been done by most studies that used this dataset because there are four missing records in the original Desharnais dataset. The Desharnais dataset, like the NASA93 dataset, contains only the year of project completion, and as such the training and test sets were formed in the same way as for the NASA93 dataset (i.e., by using the year of project completion).

According to Mann-Whitney analysis and associated boxplots, the Desharnais dataset forms a single distribution. Models were therefore built for the entire dataset along with a subset developed using a programming language termed 'Advanced Cobol' (herein referred to as the Adv.Cobol dataset). This subset is made up of 23 projects and is identified in the Desharnais dataset as "category 2" under the language attribute.

### B. Effort Estimation Model Development

In software effort estimation modelling (as in other fields) the dataset is usually split into two, forming a training set and a test set. The training set is used to develop the model and the performance of that model is then evaluated on the test set. This study follows a similar approach (see Analysis 1 and Analysis 2 in this section for model development algorithms). All models in this study are developed using the statistical package R, v.3.5.2.

All models are developed using linear regression which has enjoyed widespread use in software effort estimation studies. In order to accommodate the diverse nature of the two datasets being used in this study, especially in regard to the number of variables, specific linear regression models are applied to each dataset (or partition) as described in the respective datasets section. It should also be noted that the models developed in this study are all well-formed models. That is, the degrees of freedom are considered whereby a training set is formed only when the number of projects is at least two plus the number of explanatory variables being used for model construction. Maxwell's proposal [8] to identify influential observations using Cook's distance during model building was also adopted for this study.

---
[1] http://tinyurl.com/SEKE2020-TIME

**NASA93 Models.** In estimating effort for projects completed in a given year, equation 1, the COCOMO81 equation for effort estimation, is used for all four partitions of the NASA93 dataset.

$$effort(personmonths) = a * (KLOC)^b * (\prod EM_j) \ldots (1)$$

In order to develop a regression model, as in other COCOMO81 effort estimation studies [6] [9], equation (1) is linearized by logarithmic transformation, as indicated in equation (2).

$$\ln(effort) = \ln(a) + b * \ln(KLOC) + \ln(EM_1) + \cdots (2)$$

Backward stepwise regression is applied in order to support the inclusion or exclusion of variables, as previous studies have established that not all the effort multipliers of the NASA93 COCOMO81 format dataset are influential in model building [9].

**Desharnais Models.** Desharnais himself [7] identified the size and language attributes as those that are influential in a regression model. Kitchenham and Mendes [10] supported Desharnais' claim by proposing the use of the language attribute as a dummy variable. This approach has been adopted here for the models developed for this dataset, as shown in equation (3).

$$\ln(effort) = \ln(size) + language \ldots\ldots(3)$$

This study used the adjusted function point value as the most complete size attribute and treated the three-value language attribute as a dummy variable, with the reference dummy value being the Basic Cobol projects indicated as "1" in the Desharnais dataset. The smaller Adv.Cobol dataset only uses size as an explanatory variable in model development.

### C. Analysis Procedure

The following procedures are applied to all datasets modelled in this study.

**Analysis 1: Time-Aware Sequential Accumulation (TASA)**

1. For each dataset with timing information, select the first year in which projects were completed as the training set – if the first year of projects comprises fewer than the number of observations needed to build a well-formed model, add the next year(s) of projects, until the minimum requirement for a well-formed model is satisfied. The subsequent year of projects is then used as the test set.

2. Check for normality (Shapiro- Wilk test of normality) in the distributions of the training data– if data follow a normal distribution go to step 3 else step 2.1

2.1 Apply the appropriate transformation to make the data normal and recheck normality for verification as in step 2 above.

3. Build a regression model using the training data (where the form of the regression model will be specific to each dataset).

4. Apply the model obtained in step 3 to predict the effort values in the test set.

5. Calculate the accuracy measures (see below) for the formulae.

6. Add the test year's data to the training set; the subsequent year's data becomes the new test set.

7. Repeat steps 2 to 6 through to the estimation of the last year of projects.

**Analysis 2: Time-Aware Moving Window (TAMW)**

This algorithm applies a moving window to the dataset used in Analysis 1 thus accounting for the longevity of the projects in the training set.

1. For each dataset used in Analysis 1, drop the oldest year's projects.

2. The 'new' oldest year's projects now become the first year of projects; apply step 1 of Analysis 1.

3. Apply steps 2-6 from Analysis 1.

4. Repeat steps 3 to step 6 of Analysis 1 until the training set comprises projects from all years except the last year of projects.

5. Remove the oldest year's projects from the training set.

6. Repeat steps 1 to 5 until there is only one year of projects in the training set or until there is not enough data in the training set to build a well-formed model.

**Baseline Models.** Three baseline models are developed for each dataset/subset used in this study and their performance is compared with that of the time-aware models. The baseline models are a leave-one- out holdout (LOO – note that the 'one' in this case refers to all projects in one year rather than a single project), the mean and the median of the training set data. The mean and median effort values are calculated over the training data and become the effort estimates for the projects in the test set.

### D. Measures of Accuracy

Accuracy measures used to evaluate the performance of the effort prediction models are relative error, mean squared error and total absolute error. Note that in all three cases lower values are preferable.

*Relative Error (RE) -* The relative error is computed using the following equation:

$RE$ = variance(residuals)/variance(measured), where measured is the test data. The relative error measure accounts for the variability in data and as such it is robust to outlier data points.

*Mean Squared Error (MSE)* - MSE is defined as:

$$MSE = \frac{1}{n}\sum_{i=1}^{n}(actual - estimate)^2$$

where n is the total number of test data points, actual is the recorded effort used in developing the project and estimate is the effort predicted by the model. The MSE measures the general quality of the prediction model across all data points and accounts for projects of varying size. It can be susceptible to outliers; however, if a data set is largely free of outliers it can provide a useful indication of a model's overall accuracy.

*Total Absolute Error (TAE)* - TAE is defined as:

$$TAE = \sum_{i=1}^{n}|(actual - estimate)|$$

## IV. RESULTS

The results of applying the modelling approaches to the two datasets and their partitions are now presented. Due to space constraints we include only some of the results – the complete set of results may be found at the link specified previously.

**NASA93 Dataset.** It is evident from Table I that the accuracy measures are themselves not consistent in terms of model performance. That aside, it does seem to be feasible to build time-aware models for this dataset based on projects completion dates, as the worst model performance recorded (excluding the models with the large prediction errors) in terms of relative error is 0.26 which is quite satisfactory. Also, in just two instances the median baseline results are better than the time-aware models; in all other cases the models are better than both the mean and median baseline results.

TABLE I. NASA93 EFFORT ESTIMATION RESULTS

| | Year | Time-Aware Sequential Accumulation | | | Time-Aware Moving Window | | |
|---|---|---|---|---|---|---|---|
| | | RE | MSE | TAE | RE | MSE | TAE/AE |
| NASA82 | 1985 | 0.06 | 2136.52 | 243 | - | - | - |
| | 1986 | 1717.8 | 3.4E+07 | 18151 | - | - | - |
| | 1987 | 0.19 | 61.87 | 15 | 0.11 | 113.38 | 21 |
| C2 | 1987 | 0.26 | 73.12 | 16 | 0.26 | 73.12 | 16 |
| C5 | 1983 | - | - | 302 | - | - | 302 |
| | 1984 | 0.12 | 6256.95 | 278 | 0.12 | 6256.94 | 278 |
| | 1985 | - | - | 12 | - | - | 8 |
| | 1985* | - | - | 12 | - | - | 2 |
| SD | 1984 | 7.5E+05 | 6.1E+09 | 174344 | - | - | - |
| | 1985 | 0.02 | 199.98 | 64 | - | - | - |
| | 1986 | 849.93 | 1.7E+07 | 13996 | 1.6 | 8353.57 | 926 |
| | 1986* | 849.93 | 1.7E+07 | 13996 | 2738 | 5.4E+07 | 22513 |
| | 1987 | 0.19 | 75.92 | 17 | 0.2 | 72.17 | 16 |
| | 1987* | 0.19 | 75.92 | 17 | 0.19 | 85.47 | 17 |

\* number of additional TAMW models built for that particular year
'-' indicates no computation of a result for a specific accuracy measure

The LOO baseline results, however, is better than all the models developed for this dataset (see previous link). The highlighted results in Table I indicate large prediction errors. Manual inspection of the NASA82 and Semidetached (SD) datasets revealed that the effort multipliers of the training projects were quite different from those of the projects being estimated.

To formally gauge whether one of the time-aware models resulted in more accurate effort predictions, a two-tailed paired samples Wilcoxon test was applied. The p-value results are 1 (due mainly to the ties), 0.5839 and 0.5839 for RE, MSE and TAE, respectively. This indicates that the differences in prediction accuracy for the two models are not statistically significant. Therefore, for this dataset, we conclude that either time-aware approach could be used to estimate effort. The two time-aware models consistently included size as an explanatory variable. Beyond that, however, both the variables included in the effort estimation models and their coefficients were quite dynamic, as the models differed from one time period to another (see previous link). There was no consistent pattern as to a decrease or increase in the values of the coefficients of both model types developed for the NASA93 dataset. All the predictive models developed for the NASA93 datasets can be termed as sufficiently accurate as the Adjusted $R^2$ values fell between 0.89 and 0.98 (see previous link).

TABLE II. DESHARNAIS EFFORT ESTIMATION RESULTS

| | Year | Time-Aware Sequential Accumulation | | | Time-Aware Moving Window | | |
|---|---|---|---|---|---|---|---|
| | | RE | MSE | TAE | RE | MSE | TAE/AE |
| Desharnais | 1986 | 0.65 | 4953913.7 | 39911 | 0.67 | 4964415.7 | 40727 |
| | 1987 | 0.71 | 837535.2 | 7267 | 0.77 | 914156.4 | 7473 |
| | 1987* | 0.71 | 837535.2 | 7267 | 0.69 | 816717.2 | 7153 |
| | 1987** | 0.71 | 837535.2 | 7267 | 0.69 | 846571.1 | 8348 |
| | 1988 | 0.02 | 196984.4 | 1326 | 0.02 | 153573.1 | 1182 |
| | 1988* | 0.02 | 196984.4 | 1326 | 0.01 | 134684.7 | 1293 |
| | 1988** | 0.02 | 196984.4 | 1326 | 0.01 | 102903.6 | 1394 |
| | 1988*** | 0.02 | 196984.4 | 1326 | 0.15 | 1437442 | 3765 |
| Adv. Cobol | 1987 | 0.13 | 1393102.6 | 6515 | 0.13 | 1372071 | 6388 |
| | 1987* | 0.13 | 1393102.6 | 6515 | 0.13 | 1364820 | 6251 |
| | 1988 | - | - | 1205 | - | - | 1015 |
| | 1988* | - | - | 1205 | - | - | 1102 |
| | 1988** | - | - | 1205 | - | - | 525 |

TABLE III. COEFFICIENTS OF TIME-AWARE SEQUENTIAL ACCUMULATION MODELS - DESHARNAIS DATASET

| Dataset | Year | Intercept | Size | Lang2 | Lang3 | Adj.$R^2$ |
|---|---|---|---|---|---|---|
| Desharnais | 1986 | 5.65 | 0.50 | -0.50 | -1.66 | 0.68 |
| | 1987 | 3.78 | 0.82 | -0.04 | -1.49 | 0.74 |
| | 1988 | 3.89 | 0.80 | -0.04 | -1.44 | 0.74 |
| Adv. Cobol | 1987 | 2.66 | 1.03 | | | 0.84 |
| | 1988 | 2.62 | 1.04 | | | 0.83 |

TABLE IV. COEFFICIENTS OF TIME-AWARE MOVING WINDOW MODELS- DESHARNAIS DATASET

| Dataset | Year | Intercept | Size | Lang2 | Lang3 | Adj.$R^2$ |
|---|---|---|---|---|---|---|
| Desharnais | 1986 | 5.65 | 0.51 | -0.55 | -1.71 | 0.71 |
| | 1987 | 3.67 | 0.85 | -0.05 | -1.50 | 0.76 |
| | 1988 | 3.81 | 0.82 | -0.05 | -1.45 | 0.75 |
| | 1987* | 3.59 | 0.85 | 0.001 | -1.37 | 0.74 |
| | 1988* | 3.78 | 0.82 | -0.002 | -1.35 | 0.74 |
| | 1987** | 2.91 | 0.96 | 0.17 | -1.12 | 0.88 |
| | 1988** | 3.65 | 0.83 | 0.10 | -1.24 | 0.85 |
| | 1988*** | 4.76 | 0.62 | -0.007 | -1.06 | 0.60 |
| Adv. Cobol | 1987 | 2.74 | 1.02 | | | 0.84 |
| | 1988 | 2.96 | 0.98 | | | 0.83 |
| | 1987* | 3.01 | 0.98 | | | 0.84 |
| | 1988* | 3.37 | 0.92 | | | 0.86 |
| | 1988** | 3.32 | 0.92 | | | 0.75 |

**Desharnais Dataset.** It is evident from Table II that it is again feasible to build time-aware models for this dataset using projects completion dates, with some of the results in terms of RE reaching 0.01. The corresponding TAE results are equally satisfactory. Though the worst result, for 1986 at 40727 hours, might appear large, it equates to an average of 36 weeks per project since 28 projects were completed in 1986. There are four instances where the model results are better than their corresponding LOO(available at the previous link) baseline models (2 for MSE and 2 for RE). A two-tailed paired samples Wilcoxon test was applied to determine the superior modelling method. The p-value results are 0.6698, 0.5566 and 1 for RE, MSE and TAE, respectively, indicating that the two models are not significantly different. Therefore for this dataset, either of the time-aware approaches could be used to develop effort estimation models.

The models' explanatory variables and coefficients are consistent, as shown in Table III (TASA model) and Table VI (TAMW model). All of the models built have Adjusted $R^2$ values of between 0.60 and 0.88 and as such could be termed as reasonably accurate models.

## V. THREATS TO VALIDITY

The first threat to the validity of this study is the generalization of our results, as the datasets used are convenience sampled from the PROMISE repository. Though these datasets cannot be representative of the entire software industry they have become benchmarks datasets in software effort estimation research. The age of the datasets might also raise concern, however, these datasets are still increasingly being used in recent software effort estimation studies. Another threat to validity is due to the bias that could be introduced by considering only the completion dates, however, we had little choice as these datasets only have completion dates.

## VI. DISCUSSION AND CONCLUSIONS

The results presented for the two datasets examined here indicate that it is feasible to develop accurate effort estimation models that are also time-aware based on projects completion dates, positively answering RQ1. In most instances, the performance of the models developed for the NASA93 dataset was acceptable, with Adjusted R2 between 0.89 and 0.98 with the exception of the large errors shown in Table I.

The Adjusted R2 for the models built for the Desharnais in this study all exceeded 0.60 (better than the models built by Desharnais [7] with Adjusted $R^2$ of 0.54), most were greater than 0.70, and the highest Adjusted $R^2$ was 0.88. These results suggest that performance improvements can potentially be gained by building effort estimation models that are time-aware. The results of this study also supports Amasaki and Lokan [3] notion that it is not in all cases that time-aware models are superior. In the case of the NASA93 dataset, the LOO baseline was in fact superior to all the time aware models whilst for the Desharnais dataset, the result was mixed as the time-aware models were superior to the LOO baseline in some cases and vice-versa.

Our results regarding model stability were mixed. The variables and coefficient values for the Desharnais dataset models were generally stable, in sharp contrast to our results for the NASA93 models. The dynamic nature of the NASA93 models can be attributed to the greater heterogeneity in the NASA93 dataset – it consists of 14 different application types, developed for 5 different NASA centers, principally by a number of external vendors who may themselves have had varied development practices. The relative stability of the models built for the Desharnais dataset is somewhat surprising because this dataset was collected from ten different organizations in Canada over a period of 6 years. However, the project types and development languages used were few. This implies that it is possible that organizations working at the same time on similar projects may well have similar practices, and as such, models that are built to characterize their practice may be more homogeneous than heterogeneous. Thus, in relation to RQ2 we must conclude that *the stability of the parameters and coefficients of time-aware models largely depends on the diversity of the dataset*.

In terms of answering RQ3 as to *which of the two time-aware modelling approaches, if either, is superior in terms of accuracy*, the Wilcoxon tests indicate that there is no significant difference in performance for either of our two datasets. Our results therefore indicate that, for these two datasets, neither method is superior, and so either approach may be used to create sufficiently accurate time-aware models.

## REFERENCES

[1] M.F. Bosu and S.G. MacDonell, "A taxonomy of data quality challenges in empirical software engineering," 22nd Austral. Softw. Eng. Conf., ASWEC13, pp.97-106, 2013.
[2] M. Jorgensen and M. Shepperd, "A systematic review of software development cost estimation studies," IEEE Trans. Softw. Eng., vol. 33, no. 1, pp. 33–53, Jan. 2007.
[3] C. Lokan and E. Mendes, "Applying Moving Windows to Software Effort Estimation," Third Int. Symp. Empir. Softw. Eng. Meas., ESEM09, pp. 111–122, 2009.
[4] S. Amasaki, and C. Lokan, "An Evaluation of Selection Methods for Time- Aware Effort Estimation." In 24th Asia-Pacific Software Engineering Conference, pp. 624-629, 2017.
[5] S. G. MacDonell and M. Shepperd, "Data Accumulation and Software Effort Prediction," Proc. 2010 ACM-IEEE Int. Symp. Empir. Softw. Eng. Meas., ESEM10, pp. 31–34, 2010.
[6] B. W. Boehm, "Software Engineering Economics," IEEE Transactions on Software Engineering, vol. SE-10, no. 1. Prentice-Hall, Englewood Cliffs, NJ, Jan-1981.
[7] J.-M. Desharnais, "Statistical Analysis on the Productivity of Data Processing with Development Projects using the Function Point Technique," Université du Québec à Montréal., 1988.
[8] K. Maxwell, Applied Statistics for Software Managers. Englewood Cliffs, NJ,: Prentice-Hall, 2002.
[9] B. K. Singh, S. Tiwari, K. K. Mishra, and a. K. Misra, "Tuning of Cost Drivers by Significance Occurrences and Their Calibration with Novel Software Effort Estimation Method," Adv. Softw. Eng., vol. 2013, no. 1, pp. 1–10, 2013.
[10] B. A. Kitchenham and E. Mendes, "Why Comparative Effort Prediction Studies may be Invalid," Proc. 5th Int. Conf. Predict. Model. Softw. Eng., 2009.